\documentclass[pre,twocolumn,english,aps,prb,a4paper,floatfix]{revtex4}
\usepackage[T1]{fontenc}
\usepackage{color}
\usepackage[latin1]{inputenc}
\usepackage{graphicx}
\usepackage{bm}
\usepackage{epsfig}
\usepackage{amsmath}
\usepackage{amssymb}

\begin{document}

\title{The effect of combined roundness and polydispersity on the phase behavior of hard-rectangle fluids}
\author{Yuri Mart\'{\i}nez-Rat\'on}
 \affiliation{Grupo Interdisciplinar de Sistemas Complejos (GISC), Departamento de Matem\'aticas, Escuela Polit\'ecnica Superior, Universidad Carlos III de Madrid, Avenida de la Universidad 30, E-28911, Legan\'es, Madrid, Spain}
 \email{yuri@math.uc3m.es}
\author{Enrique Velasco}
\affiliation{Departamento de F\'{\i}sica Te\'orica de la Materia Condensada, Instituto de F\'{\i}sica de la Materia Condensada (IFIMAC) and Instituto de Ciencia de Materiales Nicol\'as Cabrera, Universidad Aut\'onoma de Madrid,
E-28049, Madrid, Spain}
\email{enrique.velasco@uam.es}

\date{\today}

\begin{abstract}
    We introduce a model for a fluid of polydisperse rounded hard rectangles where the length and width of the rectangular core are fixed, while the roundness is taken into account by the convex envelope of a disk displaced along the perimeter of the core. The diameter of the disk has a continuous polydispersity described by a Schultz distribution function. We implemented the scaled particle theory for this model with the aim to studying: (i) the effect of roundness on the phase behavior of the one-component hard-rectangle fluid, and (ii) how polydispersity affects phase transitions between isotropic, nematic and tetratic phases. We found that roundness greatly affects the tetratic phase, whose region of stability in the phase diagram strongly decreases as the roundness parameter is increased. Also the interval of aspect ratios where the tetratic-nematic and isotropic-nematic phase transitions are of first order considerably reduces with roundness, both transitions becoming weaker. Polydispersity induces strong fractionation between the coexisting phases, with the nematic phase enriched in particles of lower roundness. Finally, for high enough polydispersity and certain mean aspect ratios, the isotropic-to-nematic transition can change from second (for the one-component fluid) to first order. We also found a packing-fraction inversion phenomenon for large polydispersities: the coexisting isotropic phase has a higher packing fraction than the nematic. 
\end{abstract}

\maketitle 

\section{Introduction}

The study of entropic phase transitions in two dimensional liquid crystals is  nowadays an active line of research. This is not only because the study of orientational transitions in monolayers of anisotropic colloids or molecules (adsorbed at surfaces, interfaces or forming membranes) continues to be an interesting research topic \cite{review1,review2,Dogic}. In addition, novel techniques to synthesize hard-core interacting micro-prisms of any cross-sectional geometry, subsequently adsorbed to form monolayers, have been applied to create a plethora of effectively two-dimensional fluids of Brownian particles with several shapes \cite{Zhao5,Zhao,Zhao0,Zhao1,Zhao2,Zhao3,Zhao4}. Examples are: squares \cite{Zhao5}, rectangles \cite{Zhao}, rhombuses \cite{Zhao0}, triangles \cite{Zhao1}, hexagons \cite{Zhao2}, and kites \cite{Zhao3}. These two-dimensional fluids have been a source of fascinating phenomena including the appearance of exotic symmetries and chirality in the orientational and spatial ordering of particles. Other exotic shapes such as circular arcs \cite{Cinacchi1,Cinacchi2} have also been studied via Monte Carlo (MC) simulations, showing interesting self-assembling properties directly related to the presence of "entropic bonding", a concept recently introduced in Ref. \cite{Glotzer}. The presence of liquid-crystal and crystalline ordering was also investigated in MC simulations of two-dimensional hard regular polygons, where the number of edge-lengths play a crucial role in the symmetries of the stable liquid crystal and crystal phases \cite{Glotzer2}. 

A paradigmatic example of a two-dimensional liquid-crystal which exhibits isotropic (I), nematic (N) and tetratic (T) orientational ordering is the fluid of hard rectangles (HR) and its hard square limit. This fluid was extensively studied using the density functional theory (DFT) \cite{Schlaken, MR1,three-body}, via MC simulations \cite{Frenkel,Donev,Torres}, and even by experiments consisting of vertically shaken monolayers of granular particles of rectangular or square cross-sections \cite{Narayan,Dani,Miguel,Menon}. While particles in the I phase are not orientationally ordered, in the N or T phases the main particle axes (parallel to the major edge-length) orient along one or two equivalent directors, respectively. The  orientational distribution function $h(\phi)$, i.e. the probability density of particles axes to align at an angle $\phi$ with respect to the director, has twofold, $h(\phi)=h(\phi+\pi)$ (N phase), or fourfold, 
$h(\phi)=h(\phi+\pi/2)$ (T phase) symmetries. The T phase in the HR fluid has been shown to be stable only for aspect ratios varying from 1 (the hard-square limit, with the T phase as the only possible liquid-crystal phase) to a particular value $\kappa^*$. The value of $\kappa^*$ predicted by scaled particle theory (SPT) \cite{MR1}, a version of DFT,  and a more sophisticated DFT based on the second and third virial coefficients \cite{three-body}, are 2.21 and 3.23, respectively. Recent simulations have shown that $\kappa^*\approx 5$ \cite{Torres} while experiments on monolayers of quasi-two-dimensional granular cylinders indicate the presence of T correlations for aspect ratios as large as $\kappa_c\approx 7$ \cite{Dani,Miguel}. However, some care should be taken to compare the results from experiments on non-equilibrium dissipative granular rods to those obtained by theoretical models based on equilibrium statistical mechanics. As we have shown recently \cite{Miguel}, energy dissipation in vertically-shaken granular monolayers strongly promotes particle clusterization, resulting in a high proportion of square-like clusters made of rectangles joined side by side, which in turn induces the formation of T textures as stationary states. 

On the other hand, confined monolayers of cylinders respond to geometrical frustration much in the same way as equilibrium liquid crystals, i.e. by creating topological defects that restore the global symmetry of the system. The number and topological charge of these defects seem to follow the rules of topology. Also, when particles are confined in annular geometries, a complicated pattern arises in the orientational-ordering field, with domain walls that separate regions of smectic and T ordering and additional topological defects \cite{Ariel}. While topology predicts that no defects should be excited in this case, the small size of the system compared with the particle length probably explains the formation of this complex pattern, although non-equilibrium effects cannot be discarded as an origin. Colloidal monolayers also exhibit the presence of T-like disclination defects in the smectic textures when confined inside cavities of different shapes \cite{Lowen1,Lowen2}. These similarities between dissipative and equilibrium monolayers point to the preponderant role of entropic interactions as the main mechanism dictating the symmetries of both systems when frustrated by confinement. 

The T phase can be stabilized by other geometrical shapes such as rhombuses \cite{Torres} and kites \cite{Zhao3,MR4} of particular shapes and ratios between their characteristic lengths. Indeed its stability region in the phase diagram seems to be very sensitive to these ratios and, what can be more important, to the roundness of the particle corners. It was recently shown by MC simulations that a fluid of hard rounded squares does not exhibit a T phase for high enough roundness of the corners, with the I phase directly undergoing a transition to a crystalline phase \cite{Escobedo}. This result explained why the T phase was not found in recent experiments on rounded squares \cite{Zhao5}: it is certainly difficult to design an experimental procedure to obtain microparticles with perfect right corners.  There are two ways to implement the presence of nonzero curvature in the particle boundaries: (i) to take into account the change of curvature by defining the particle as a superellipse \cite{Torres,Varga1,Varga2} with an exponential parameter ranging from 1 (the rhomboidal shape), 2 (the elliptical shape), and finally the infinite limit (for the case of rectangles), or (ii) to consider a fixed core defined by straight lines, adding the convex envelope that results when a disk of some particular diameter slides along the boundary of this core \cite{Escobedo}. 

Here we will use the second recipe, with a fixed core of rectangular shape, defining in such a way a fluid of hard rounded rectangles (HRR).  Two  main studies have been carried out. In the first we study the effect of roundness  (measured through a roundness parameter) on the stability of the T phase as compared with the HR fluid. Using scaled particle theory (SPT) \cite{Cotter,Lasher,Barboy}, several phase diagrams for different values of the roundness parameters have been calculated, which allowed to trace out the stability boundaries of the different phases and their changes as a function of the roundness parameter. The second study deals with the effect of polydispersity in the phase behavior of HRR. This point is motivated by the fact that some polydispersity in sizes/shapes is always present in the experimental systems. As will be shown later, the main effects of particle roundness on the one-component fluid of HRR are: (i) the strong destabilization of the T phase; the stability region of the T phase in the packing fraction-aspect ratio plane is considerably shrinked as roundness increases, (ii) the interval of aspect ratios where the I-N and T-N transitions are of first order is strongly reduced, and as a consequence both transitions become weaker. As regards the effect of polydispersity we find that, for certain aspect ratios and high enough mean roundness and polydispersity coefficient, the I-N transition for certain mean aspect ratios becomes of first order despite being of second order in the one-component fluid. Also when the fluid exhibits a first-order I-N or T-N transition, and for high enough polydispersity, the coexisting phases exhibit a packing fraction inversion due to the fractionation effect: the coexisting I or T phases are enriched in particles of higher roundness (or lower aspect ratios), while the N phase is more populated by species of higher aspect ratios. As a consequence, the coexisting I or T phases can have a lower packing fraction as compared to that of the coexisting N.  We should mention that previous MC simulation studies on monodisperse hard rods in 2D showed that a quasilong-range ordered N phase exhibits a transition to the I phase via a Kosterlitz-Thouless disclination unbinding type mechanism rather than being of first order \cite{Frenkel2,Dijkstra}. However recent studies have shown that, for particular types of particle interactions, the I-N transition becomes of first order in 2D \cite{Enter,Vink}. Finally, recent experiments found that quasi-monolayers of magnetic nanorods confined between adjacent layers of a lamellar phase exhibit a first order I-N transition \cite{Constantin}. 

The theoretical DFT study of continuous polydisperse fluids  of anisotropic particles represents a challenge because the density profile depends not only on the particle orientational degrees of freedom. It also incorporates the distribution of the polydisperse variable, which complicates the numerical procedure necessary to calculate phase coexistence. To deal with this problem, some simplifications were made in the past to study the effect of polydipersity on the phase behavior of freely-rotating hard polydisperse rods. One of these simplifications involves using the Onsager-DFT of hard spherocylinders in the hard-needle limit, and implementing the spherical harmonics expansion of the excluded volume up to second order, together with the use of the moment theory \cite{Sollich1a,Sollich1b} to render the calculations feasible \cite{Sollich2}. An alternative approach is to discretize the orientational degrees of freedom, as in the Zwanzig approximation, and use the Fundamental Measure DFT for hard boardlike particles which correctly describes not only two-body, but also three-body correlations \cite{Cuesta}. As we show here the present model has the advantage that the orientational degrees of freedom and the polydisperse variable (the diameter of the disk causing particle roundness) are decoupled, making the theoretical treatment of the polydisperse fluid as easy as its one-component counterpart. This property can be used in future developments to study the combined effect of confinement and polydispersity on the structural properties of a two-dimensional liquid-crystal fluid.     
The article is organized as follows. Sec. \ref{model} is devoted to the definition of the model and the presentation of the theoretical tools used for the calculation of phase diagrams. Special attention is paid to the coexistence calculation  formalism (Sec. \ref{coex_sec}), the definition of the polydisperse distribution function used in the study (Sec. \ref{parent_sec}), and the implementation of the bifurcation analysis to calculate the second-order phase transition curves (Sec. \ref{bifurcation}). The results are divided in two parts: In Sec. \ref{one-component} we present the results for the one-component fluid (zeroth polydispersity), while in Sec. \ref{polydisperse_sec} we describe the results regarding the effect of polydispersity on the phase behavior of HRR. In Sec. \ref{crystal} we describe an approximate procedure to account for the effect of roundness on the instability of the T phase with respect to crystallization. Finally some conclusions and discussions are summarized in Sec. \ref{conclusions}. We relegate to Sec. \ref{app} the details for the numerical calculations of shadow and cloud coexistence curves in the polydisperse HRR fluid.

\section{Model and Theory}
\label{model}

\begin{figure}
	\epsfig{file=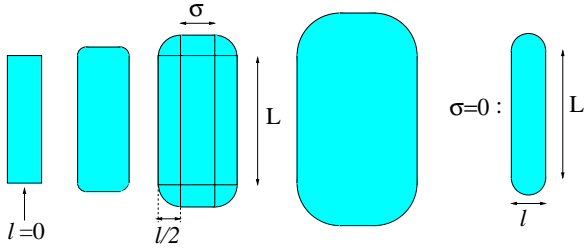,width=3.in}
	\caption{Sketches of rounded rectangles where the fixed core lengths $L$ and $\sigma$ 
	and the polydisperse disk length $l$ of particle are indicated. Note that the limiting 
	case $l=0$ corresponds to a rectangular particle, while for 
	$l\neq 0$ and $\sigma=0$ one obtains a  
	discorectangle.}
	\label{fig1}
\end{figure}

Our model consists of a polydisperse mixture of hard rounded rectangles (HRR). A particle is defined by a fixed rectangular core of length $L$ and width $\sigma$ ($L\geq \sigma$) plus a polydisperse coating obtained by making the center of a disk of diameter $l$ to slide over the perimeter of the rectangular core (see Fig. \ref{fig1} 
for a sketch of the particle geometry). While the core sizes $L$ and $\sigma$ are fixed, 
the diameter $l$ is a polydisperse variable with a value ranging from 0 to $\infty$. 
Note that the cases $l=0$ and $\sigma=0$ constitute the limiting cases of hard rectangular and discorectangular particles, respectively. 
In the following we define the main axis of a particle to be parallel to its length $L$.

Let us consider two such particles with different values ($l$ and $l'$) of the disk diameters. The excluded area between these particles, as a function of their
relative angle $\Delta \phi=\phi-\phi'$, can be computed as 
\begin{eqnarray}
	&&A_{\rm excl}(\Delta \phi,l,l')=\left(L^2+\sigma^2\right)
	|\sin \Delta \phi|+2L\sigma|\cos \Delta\phi|\nonumber\\
	&&+\left(L+\sigma\right)\left(l+l'\right)+\frac{\pi}{2}ll'+a(l)+a(l'), 
	\label{first}
\end{eqnarray}
where the particle area, $a(l)$,  is  
\begin{eqnarray}
	a(l)=L\sigma+(L+\sigma)l+\frac{\pi}{4}l^2.
	\label{area}
\end{eqnarray}
From the excluded area we define the geometric function from which the scaled particle 
theory is constructed: 
\begin{eqnarray}
	A_{\rm spt}(\Delta \phi,l,l')=\frac{1}{2}\left[ A_{\rm excl}(\Delta\phi,l,l')-a(l)-a(l')\right]. 
	\label{spt0}
\end{eqnarray}
The polydisperse mixture of HRRs is characterized by its number density distribution function, 
$\rho(l,\phi)$, a probability density in the variables $l$ and 
$\phi$. This function describes the probability to find a particle with disk size $l$ oriented 
with an angle $\phi$ with respect to a fixed reference frame. Thus we have  
$\displaystyle{\int_0^{\infty} d l\int_0^{2\pi} d\phi \rho(l,\phi)=\frac{N}{A}}$, with $N$ and $A$ the total number of particles 
and the area of the system. From $\rho(l,\phi)$ we can define its $i$th moment $m_i(\phi)$ with respect to $l$ and the integral of this moment with respect to $\phi$ as 
\begin{eqnarray}
	m_i(\phi)\equiv \int_0^{\infty} dl \ l^i \rho(l,\phi), 
	\quad m_i\equiv \int_0^{2\pi} d\phi \ m_i(\phi). \label{moments}
\end{eqnarray}
Note that $m_0=N/A$, the total number density. To facilitate numerical calculations we will use the Fourier expansion of the function $m_0(\phi)$,
\begin{eqnarray}
	m_0(\phi)=\frac{1}{2\pi}\left[m_0+\sum_{k\geq 1} m_0^{(k)} \cos(2k\phi)\right],
	\label{expansion}
\end{eqnarray}
with $\{m_0^{(k)}\}$ the Fourier amplitudes. From $m_0(\phi)$ the orientational distribution function is simply $h(\phi)=m_0(\phi)/m_0$, while the order parameters describing the orientational ordering are
\begin{eqnarray}
	Q_{2n}\equiv \int_0^{2\pi} d\phi h(\phi)\cos(2n\phi)=
	\frac{m_0^{(n)}}{2m_0}, \quad n=1,2
\end{eqnarray}
For uniaxial N orientational symmetry we have $Q_2\neq 0$,  while T symmetry is characterized by $Q_2=0$ and $Q_4\neq 0$. 

The double average of the function 
$A_{\rm spt}(\phi-\phi',l,l')$ with respect to $\rho(l,\phi)$ and $\rho(l',\phi')$ gives
\begin{eqnarray}
	&&\langle\langle A_{\rm spt}(\Delta\phi,l,l')\rangle\rangle_{\rho(l,\phi)}
	\equiv \int_0^{\infty} dl \int_0^{\infty} dl'\int_0^{2\pi} d\phi 
	\nonumber\\ 
	&& \times \int_0^{2\pi} d\phi' 
	\rho(l,\phi)
	\rho(l,\phi') A_{\rm spt}(\phi-\phi',l,l')\nonumber\\
	&&= \frac{\left(L^2+\sigma^2\right)}{2}\langle\langle|\sin(\Delta\phi)|
	\rangle\rangle_{m_0(\phi)}\nonumber\\
	&&+L\sigma\langle\langle|\cos(\Delta\phi)|\rangle\rangle_{m_0(\phi)}
	+(L+\sigma)m_0m_1+\frac{\pi}{4}m_1^2, \nonumber\\
	\label{spt}
\end{eqnarray}
where we have used the shorthand notation  
\begin{eqnarray}
	&&\langle\langle g(\Delta\phi)\rangle\rangle_{m_0(\phi)}=
	\int_0^{2\pi}d\phi\int_0^{2\pi}d\phi'm_0(\phi)m_0(\phi')\nonumber\\
	&&\times g(\phi-\phi'). 
\end{eqnarray}
Inserting the Fourier expansion (\ref{expansion}) into Eq. (\ref{spt}) we obtain 
\begin{eqnarray}
	&&\langle\langle A_{\rm spt}(\Delta\phi,l,l')\rangle\rangle_{\rho(l,\phi)}
	=\frac{g_0m_0^2-\frac{1}{2}\sum_{k\geq 1}
	g_k\left(m_0^{(k)}\right)^2}{\pi}\nonumber\\
	&&+(L+\sigma)m_0m_1+\frac{\pi}{4}m_1^2, 
		\label{double}
\end{eqnarray}
with
\begin{eqnarray}
    g_k\equiv \frac{\left(L+(-1)^k\sigma\right)^2}{4k^2-1}
	\label{double1}
\end{eqnarray}
Another important quantity of the polydisperse mixture is the total packing fraction:
\begin{eqnarray}
	&&\eta=\int_0^{\infty}dl\int_0^{2\pi}d\phi \rho(l,\phi)a(l)=m_0L\sigma+m_1\left(L+\sigma\right)\nonumber\\
	&&+\frac{\pi}{4}m_2, \label{def_packing}
\end{eqnarray}
In this expression we used (\ref{area}) for the particle area and (\ref{moments}) for the integrated moments of $\rho(l,\sigma)$. Note that $\eta$ depends not only on the the zeroth and first moments $m_0$ and $m_1$, as does the double average of the scaled-particle area (\ref{double}), but also on the second moment $m_2$.

With these definitions, the excess part of the free-energy density, according to the SPT \cite{three-body}, can be calculated (in thermal units), as
\begin{eqnarray}
	&&\Phi_{\rm ex}[\rho(l,\phi)]\equiv \frac{\beta {\cal F}_{\rm exc}[\rho(l,\phi)]}
	{A}=-m_0\log(1-\eta)\nonumber\\
	&&+\frac{\langle\langle A_{\rm spt}(\Delta\phi,l,l')\rangle 
	\rangle_{\rho(l,\phi)}}{1-\eta},
\end{eqnarray}
while the ideal part is, as usual,
\begin{eqnarray}
	&&\Phi_{\rm id}[\rho(l,\phi)]\equiv \frac{\beta{\cal F}_{\rm id}[\rho(l,\phi)]}{A}\nonumber\\&&=
	\int_0^{\infty} dl\int_0^{2\pi}\rho(l,\phi)\left[\log \rho(l,\phi)-1\right].
\end{eqnarray}
In the above expressions, 
$\beta {\cal F}_{\rm id,exc}[\rho(l,\phi)]$ are the ideal and the excess parts 
of the free-energy density functional, scaled with the factor
$\beta=\left(k_B T\right)^{-1}$ (note that thermal area inside the logarithm of the ideal part has
been dropped).

\subsection{Coexistence calculations}
\label{coex_sec}

Now we calculate the two-phase coexistence between a phase that
occupies a fraction $1-\epsilon$ of the total area (the cloud phase), and another phase that occupies a vanishingly small fraction of the area, $\epsilon\ll 1$ (the shadow phase), with coexisting density distributions $\rho_{\rm c}(l,\phi)$ and $\rho_{\rm s}(l,\phi)$, respectively. 

Let us obtain the equations that govern this coexistence.
Mass conservation, expressed by the lever rule, states that the sum of the two density
distributions, integrated over the angle $\phi$ and each multiplied by its respective 
area occupancy, $1-\epsilon$ or $\epsilon$, is a conserved quantity. This is equal to the distribution function of the parent phase, $\rho_0(l)\equiv \rho_0 f(l)$, where $\rho_0$ is the total number density of 
the system, $\rho_0=N/A$, while $f(l)$ is a fixed probability disk-diameter distribution function. 
The lever rule is then
\begin{eqnarray}
	\rho_0f(l)=(1-\epsilon) \int_0^{2\pi} d\phi \rho_c(l,\phi)+\epsilon \int_0^{2\pi} d\phi \rho_s(l,\phi).
	\label{level}
\end{eqnarray}
Minimizing the total free-energy density $\Phi[\rho(l,\phi)]=\Phi_{\rm id}[\rho(l,\phi)]+
\Phi_{\rm ex}[\rho(l,\phi)]$  with respect 
to $\rho_{c,s}(l,\phi)$, and using the lever rule (\ref{level}) and the integral expression (\ref{spt}) for the averaged scaled particle area, we obtain
\begin{eqnarray}
	\rho_{c,s}(l,\phi)=e^{\beta \mu_0(l)} e^{-c_1^{(c,s)}(l,\phi)}, \label{insert}
\end{eqnarray}
where the Lagrange multiplier $\beta\mu_0(l)$, necessary to satisfy the constraint (\ref{level}), 
is just the scaled chemical potential of the species with disk-diameter value $l$. In the above
we have used the notation $c_1^{(\alpha)}(l,\phi)$ for the first functional derivative of the
excess free energy:
\begin{eqnarray}
	&&c_1^{(\alpha)}(l,\phi)=\frac{\delta \Phi_{\rm ex}[\rho_{\alpha}(l,\phi)]}
	{\delta \rho_{\alpha}(l,\phi)}\nonumber\\
	&&=-\log\left(1-\eta_{\alpha}\right)+\frac{1}{1-\eta_{\alpha}}
	\times \frac{\delta \langle\langle A_{\rm spt}(\Delta\phi,l,l')\rangle\rangle
	_{\rho_{\alpha}(l,\phi)}}{\delta\rho_{\alpha}(l,\phi)}\nonumber\\
	&&+\beta p_{\alpha} a(l),
	\label{c1}
\end{eqnarray}
where 
\begin{eqnarray}
\beta p_{\alpha}=\frac{m_0^{(\alpha)}}{1-\eta_{\alpha}}
	+\frac{\langle\langle A_{\rm spt}(\Delta\phi,l,l')\rangle\rangle
	_{\rho_{\alpha}(l,\phi)}}{(1-\eta_{\alpha})^2}, \ \alpha=c,s,
\end{eqnarray}
is the pressure of the coexisting $\alpha$-phase. Using the definition 
(\ref{spt}) and the Fourier expansion (\ref{expansion}), we explicitly find the first functional 
derivative of the scaled particle area:
\begin{eqnarray}
	&&\frac{\delta \langle\langle A_{\rm spt}(\Delta\phi,l,l')\rangle\rangle
	_{\rho_{\alpha}(l,\phi)}}{\delta\rho_{\alpha}(l,\phi)}\nonumber\\
	&&=\frac{2}{\pi}\left[g_0m_0^{(\alpha)}
	-\sum_{k\geq 1} g_km_0^{(k,\alpha)}
	\cos(2k\phi)\right]\nonumber\\
	&&+\left(L+\sigma\right)m_1^{(\alpha)}+\left[\left(L+\sigma\right)m_0^{(\alpha)}+
	\frac{\pi}{2}m_1^{(\alpha)}\right]l. \label{delta}
\end{eqnarray}
The Lagrange multiplier $\mu_0(l)$ can be found by inserting (\ref{insert}) into the lever rule 
(\ref{level}), and taking the limit $\epsilon\to 0$, which allows us to 
obtain expressions for the coexisting cloud and shadow densities:
\begin{eqnarray}
	&&\rho_{\alpha}(l,\phi)=\rho_0 f(l) \frac{e^{-c_1^{(\alpha)}(l,\phi)}}{\displaystyle\int_0^{2\pi} d\phi'	e^{-c_1^{(c)}(l,\phi')}}, \ \alpha=c,s.
	\label{uno}
\end{eqnarray}
Finally, multiplying Eqns. (\ref{uno}) by $l^i\cos(2k\phi)$, and integrating over $\phi$ and $l$, we find 
\begin{eqnarray}
	&&m_i^{(k,\alpha)}=\frac{2\rho_0}{1+\delta_{k0}}
	\int_0^{\infty} dl l^i f(l)\nonumber\\
	&&\times \frac{\displaystyle\int_0^{2\pi} d\phi \cos(2k\phi)
	e^{-c_1^{(\alpha)}(l,\phi)}}{\displaystyle\int_0^{2\pi}d\phi' e^{-c_1^{(c)}(l,\phi')}}, 
	\ \alpha=c,s; \ i=0,1,2,\nonumber\\
	\label{todas}
\end{eqnarray}
where $m_i^{(k,\alpha)}$ is defined as the $k$th-order Fourier coefficient of the moment (\ref{moments}).

For the cloud-coexisting phase and $k=0$, we obtain  
$m_i^{(0,c)}=m_i^{(c)}=\rho_0\int_0^{\infty} dl f(l) l^i=\rho_0\langle l^i\rangle_{f(l)}$, which 
coincides with the $i$th-moment of the parent distribution function. Moreover if 
the cloud-phase is I, we have $m_i^{(k,c)}=0$ $\forall $ $k\geq 1$.
We have solved a subset of Eqns. (\ref{todas}), together with the pressure equality, $p^{(c)}=p^{(s)}$, between cloud and shadow phases, to find the set of moments 
$\{m_i^{(k,\alpha)}\}$ in both coexisting phases and at the parent 
number density $\rho_0$. As will be shown in Sec. \ref{app}, we need to solve 
a total number of equations less than that in (\ref{todas}), which is a direct consequence of the 
peculiar form of the spt-area (\ref{spt0}). 

\subsection{The polydisperse probability parent distribution function}
\label{parent_sec}

In the present study we use a Schultz distribution to describe the polydispersity in $l$ in the parent phase:
\begin{eqnarray}
	f(l)=\frac{(\nu+1)^{\nu+1}}{l_0\Gamma(\nu+1)}\left(\frac{l}{l_0}\right)^{\nu}
	e^{-(\nu+1)l/l_0}, \label{parent}
\end{eqnarray}
where $l_0=\langle l\rangle_{f(l)}$ is the mean value. $\Gamma(x)$ is the Gamma function.
The above expression fulfills the normalization condition $\int_0^{\infty} dl f(l)=1$.
The parameter $\nu\in[0,\infty]$ is related to the mean square deviation by
\begin{eqnarray}
	s\equiv\sqrt{\frac{\langle l^2\rangle_{f(l)}}{\langle l\rangle_{f(l)}^2}-1}=\frac{1}
	{\sqrt{\nu+1}}.
\end{eqnarray}
The parameter $s\in[0,1]$ is used as a measure of polydispersity. 
Note that the second moment is given in terms of $s$ by $\langle l^2\rangle_{f(l)}=l_0^2(1+s^2)$.

To measure the fractionation between the coexisting phases, we will use 
the mean disk size with respect to the distribution function of the shadow phase:
\begin{eqnarray}
	f^{(s)}(l)\equiv \frac{1}{m_0^{(s)}}\int_0^{2\pi} d\phi \rho^{(s)}(l,\phi),
\end{eqnarray}
and scaled with the mean size $l_0$ of the cloud phase. The result is  
\begin{eqnarray}
	\frac{\langle l\rangle_{f^{(s)}(l)}}{l_0}=
	\frac{m_1^{(s)}}{m_0^{(s)} l_0}.\nonumber
\end{eqnarray}
Obviously this magnitude is unity for the cloud phase.

\subsection{I-N and I-T bifurcations}
\label{bifurcation}
The packing fraction at bifurcation between the I phase 
and the orientationally ordered N or T phases can be obtained from a bifurcation analysis
(see the Appendix \ref{app}). This value gives the exact location of the second-order transition, and corresponds to the 
spinodal instability of the I phase in the case of a first-order transition.  
Here we express the result given by Eqn. (\ref{bifurca}) for the packing fractions at bifurcation as a function 
of new parameters characterizing particle geometry. We define 
the mean roundness parameter $\theta$ and the mean aspect ratio $\kappa$ of the particle as 
\begin{eqnarray}
	\theta\equiv \frac{l_0}{\sigma+l_0}, \quad 
	\kappa\equiv \frac{L+l_0}{\sigma+l_0}. \label{coeff}
\end{eqnarray}
We can see that $\theta=0$ for the perfect rectangular particle ($l_0=0$) 
and $\theta\to 1$ for a very large roundness, $l_0\gg \sigma$. 
In terms of these new variables, the I-N ($k=1$) and I-T ($k=2$) bifurcation point
takes place at packing fractions given by
\begin{eqnarray}
	\eta_1=\left\{1+\frac{2(\kappa-1)^2}{3\pi
	\left[\kappa-\theta^2\left(1-\pi(1+s^2)/4\right)\right]}\right\}^{-1}, \label{la_eta1}\\
	\eta_2=\left\{1+\frac{2(\kappa+1-2\theta)^2}{15\pi
	\left[\kappa-\theta^2\left(1-\pi(1+s^2)/4\right)\right]}\right\}^{-1}. \label{la_eta2}
\end{eqnarray}
The crossover aspect ratio $\kappa_c$ when the I-N and I-T bifurcation curves coincide, i.e. for 
$\eta_1(\kappa)=\eta_2(\kappa)$, is a linear function of $\theta$:
\begin{eqnarray}
\kappa_c=\kappa^*-(\kappa^*-1)\theta,\quad \kappa^*=
\frac{3+\sqrt{5}}{2},
\end{eqnarray}
where $\kappa^*\approx 2.618$ is the cross-over aspect ratio corresponding 
to hard rectangles ($\theta=0$). For $\kappa<\kappa_c$ ($\kappa>\kappa_c$) the stable phase above the 
bifurcation curve is T (N). We can see that $\kappa_c$ decreases with $\theta$ indicating that 
the roundness destabilizes the T phase. The packing fraction value at $\kappa_c$ is 
\begin{eqnarray}
	&&\eta_c=\eta_1(\kappa_c)=\eta_2(\kappa_c)\nonumber\\
	&&=\left\{1+\frac{2\kappa^*(1-\theta)^2}
	{3\pi\left[(\kappa^*+\theta)(1-\theta)+\pi\theta^2(1+s^2)/4\right]}
	\right\}^{-1}.
	\label{packing_I-N}
\end{eqnarray}
Also, for polydisperse rectangles close to the hard-square shape 
($\kappa=1$), the I phase bifurcates to the T phase at packing fraction 
\begin{eqnarray}
	\eta_2(1)=\left\{1+\frac{8(1-\theta)^2}{15\pi\left[1-\theta^2\left(1-\pi(1+s^2)/4\right)\right]}\right\}^{-1}.
\end{eqnarray}

\begin{figure}
	\epsfig{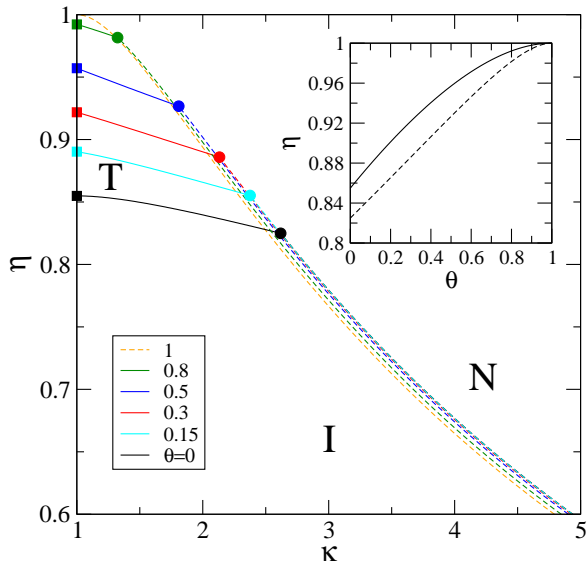}
	\caption{I-N ($\eta_1(\kappa)$ with dashed lines) and I-T ($\eta_2(\kappa)$ with solid lines) bifurcation 
	curves in the one-component fluid of rounded hard rectangles. The solid circles indicate the 
	location of the (I-N)--(I-T) crossover points at $\kappa_c$, while solid squares correspond to the I-T bifurcation packing fractions for hard squares. The inset show 
	the crossover packing fraction value $\eta_c=\eta_i(\kappa_c)$, $i=1,2$ 
	(dashed line) and the packing fraction at the I-T bifurcation, $\eta_2(1)$, for hard squares, both as a function 
	of the roundness $\theta$.}
	\label{fig2}
\end{figure}

We firstly analyze the case $s=0$, the one-component fluid. In Fig. \ref{fig2} we plot the functions $\eta_1(\kappa)$ 
(dashed lines) and $\eta_2(\kappa)$ (solid lines)
for values of $\theta$ belonging to the set $\{0,\ 0.15,\ 0.3,\ 0.5,\ 0.8,\ 1\}$. We can see that the N phase  
stabilizes at lower densities as $\kappa>\kappa_c$ increases, while for $\kappa<\kappa_c$ 
the I phase bifurcates to the T phase. Note that the function $\eta_1(\kappa)$ for a fixed value of $\kappa$ is 
a decreasing function of $\theta$ although the variation is rather small. This in turn means that the roundness 
keeps approximately the same I-N bifurcation value. From the 
figure we confirm that the 
region of T phase stability strongly decreases with $\theta$ and disappears altogether for $\theta=1$ (hard discorectangles): Not only the value of $\kappa_c$ decreases with $\theta$ but also the packing fraction 
$\eta_2(\kappa)$ dramatically increases with $\theta$. In the inset the functions 
$\eta_i(\kappa_c)$, $i=1,2$ (dashed) and $\eta_2(1)$ (solid) are plotted as a function of $\theta$, both being monotonically increasing functions, a direct consequence of the destabilizing effect of roundness on the T phase. 

\begin{figure}
	\epsfig{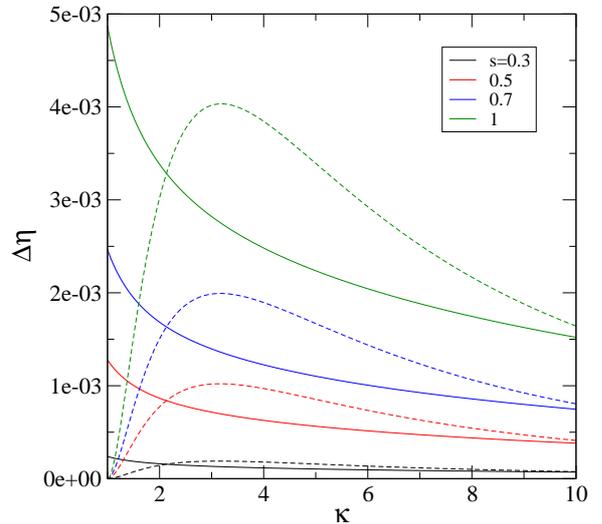}
	\caption{The difference, $\Delta \eta(\kappa)=\eta_i(\kappa;s)-\eta_i(\kappa;0)$ ($i=1$: dashed, 
	and $i=2$: solid), between the bifurcation packing fractions of polydisperse and one-component 
	fluid for a fixed value of the roundness $\theta=0.3$ and different values of the polydisperse 
	coefficient as they are shown.}
	\label{fig3}
\end{figure}

For a fixed roundness $\theta$ the polydispersity has the effect of increasing the packing 
fractions at I-N and I-T bifurcations, which can be seen in Fig. \ref{fig3} where the difference, 
$\Delta \eta(\kappa)\equiv\eta_i(\kappa;s)-\eta_i(\kappa;0)$ ($i=1,2$), between the 
bifurcation packing fractions of polydisperse and one-component fluids for the selected set of 
polydisperse coefficients $s=\{0.3,\ 0.5,\ 0.7,\ 1\}$ are plotted. From the figure we conclude that this effect is rather small.

As we will show in Sec. \ref{one-component}, the I-N transition is of first order for $\kappa_c\alt\kappa<\kappa_t$, with 
$\kappa_t$ the aspect ratio value of the I-N tricritical point, its value strongly depending on $\theta$. For 
this range of $\kappa$ we should bear in mind that the packing fractions at which the N phase begins 
to be stable do not coincide with the bifurcation values calculated here. Also when polydispersity is large enough,
the shadow and cloud curves have a large coexisting gap, strongly deviating from the bifurcation curves.   

\section{The one-component fluid}
\label{one-component}

\begin{figure*}
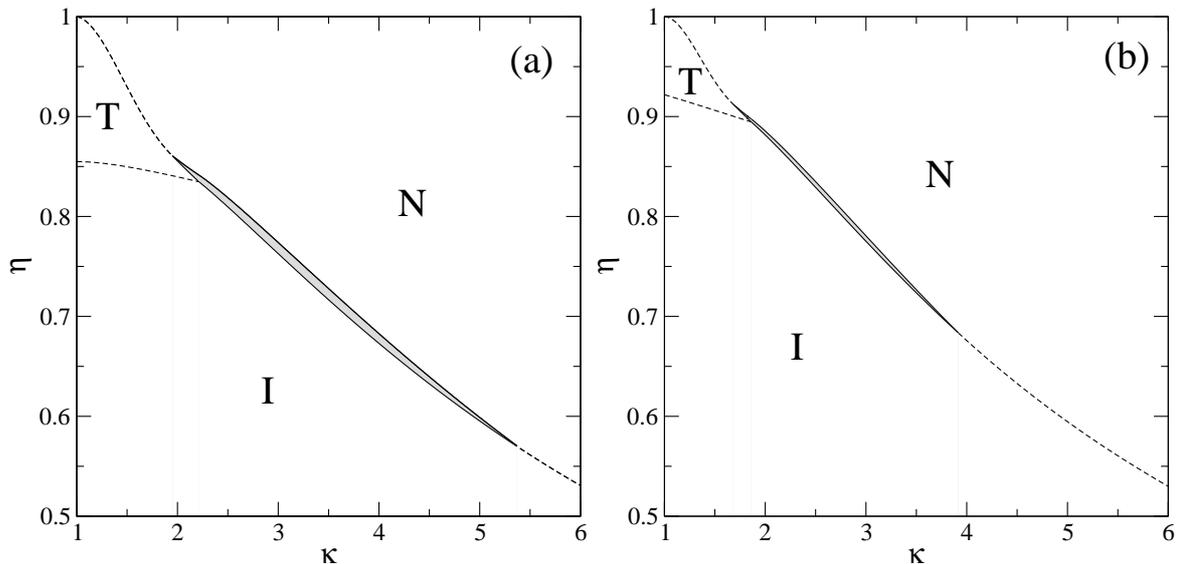

	\epsfig{file=fig4a.eps,width=3.in}
	\epsfig{file=fig4b.eps,width=3.in}
	\caption{Phase diagrams packing fraction, $\eta$, vs. aspect ratio, $\kappa$, of 
	the one-component HR (a) and RHR (b) fluids. The roundness parameter for the later 
	is $\theta=0.3$. The regions of stability of I, T and N phases are correspondingly labeled. With solid and dashed lines we show second and first order phase transition. For the later the coexistence gaps are shaded in grey. The regions of stability of the I, N and T phases are labeled.}
	\label{fig4}
\end{figure*}

In this section we present results for the phase behavior of the one-component 
fluid of HRR. The polydisperse coefficient is set as $s=0$ in the system of Eqns. (\ref{looking})-(\ref{looking3}), which are solved numerically, together with the pressure equality condition, to calculate the coexistence between the I or T phases and the N phase. In the case of a second-order transition the expressions (\ref{la_eta1}) and (\ref{la_eta2}) are used to compute the packing fraction at the I-N and I-T bifurcations, respectively, or else Eqn. (\ref{linear}) to find the T-N bifurcation numerically. Fig. \ref{fig4}(a) shows the phase diagram of the HR fluid ($\theta=0$) already obtained in Ref. \cite{MR1}, which is plotted here for the sake of comparison. In panel (b) the phase diagram of HRR with roundness $\theta=0.3$ is shown. The main differences between the phase diagrams are: (i) The region of stability of the T phase of HRR shrinks considerably (the second-order I-T bifurcation moves to higher packing fraction substantially, while the end critical point (the point at which the I-T second-order line and the I binodal of the I-N coexistence meet) moves to smaller aspect ratios. This result is in agreement with the evolution of the bifurcation curves as $\theta$ increases, a point already discussed in Sec. \ref{bifurcation}. (ii) The range of particle aspect ratios located between the T-N and I-N tricritical points (the left and right ends of the continuous lines) is considerably smaller as compared to the HR fluid, i.e. the interval in $\kappa$ for which the I-N transition is of first order strongly reduces. Note that the aspect ratio at the I-N tricritical point is smaller than its HR counterpart. Not only that: also the coexistence gap (compare the grey shaded regions inside the coexisting binodals in both panels) is much smaller, indicating a weaker first-order transition. We can conclude that particle roundness strongly destabilizes the T phase, making the fluid prone to exhibiting continuous phase transitions. These features become more pronounced as the roundness parameter $\theta$ increases. In the limit $\theta\to 1$ we obtain the phase diagram of hard discorectangles, with a simple second-order I-N transition for any aspect ratio.

\begin{figure}
	\epsfig{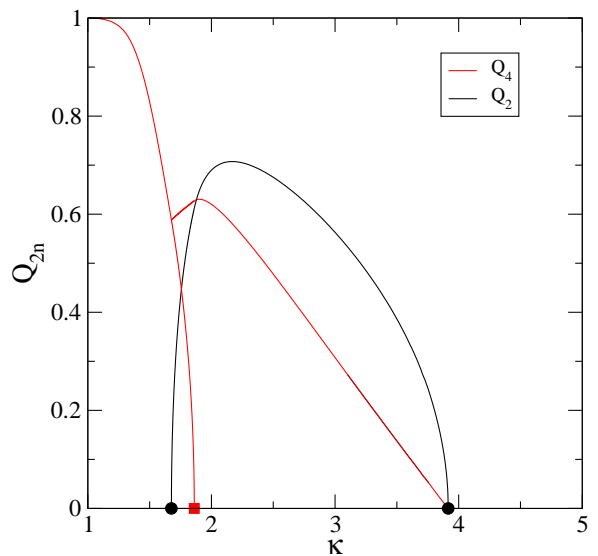}
	\caption{N ($Q_2$) and T ($Q_4$) order parameters along the I-N, I-T coexisting binodals 
	and also along the T-N second order transition corresponding to a fluid of HRR with 
	roundness $\theta=0.3$. Solid circles indicate the positions of 
	the T-N (left) and I-N (right) tricritical points while the solid square is at the 
	critical end-point where the I-T second order line meets the T binodal of the T-N coexistence (at the left of the point) and the I binodal of the I-N transition (at the right). See phase diagram of Fig. \ref{fig4} (b).}
	\label{fig5}
\end{figure}

We now proceed to describe the orientational ordering along the coexisting and second-order curves of the phase diagram in Fig. (\ref{fig4})(b). In Fig. \ref{fig5} the order parameters $Q_2$ and $Q_4$ along these curves are shown. $Q_2$ is different from zero between the T-N (at $\kappa=1.68$) and I-N (at $\kappa=3.91$) tricritical points, shown with solid circles in the figure, and exhibits a maximum around  $\kappa\approx 2.17$ where the I-N transition is strongly of first order. At the left of $\kappa\approx 1.86$ (the location of the critical end-point where the second order I-T transition and the first-order I-N transition meet, indicated by a solid square), the I-N transition continues as a T-N transition with the order parameter $Q_4$ of the T phase increasing as $\kappa$ decreases up, to the intersection with $Q_4$ of the N phase at the T-N tricritical point. For still lower values of $\kappa$ the T-N transition is always of second order and the T order parameter $Q_4$ increases along the T-N bifurcation curve up to a value of unity at $\kappa=1$.

\begin{figure}
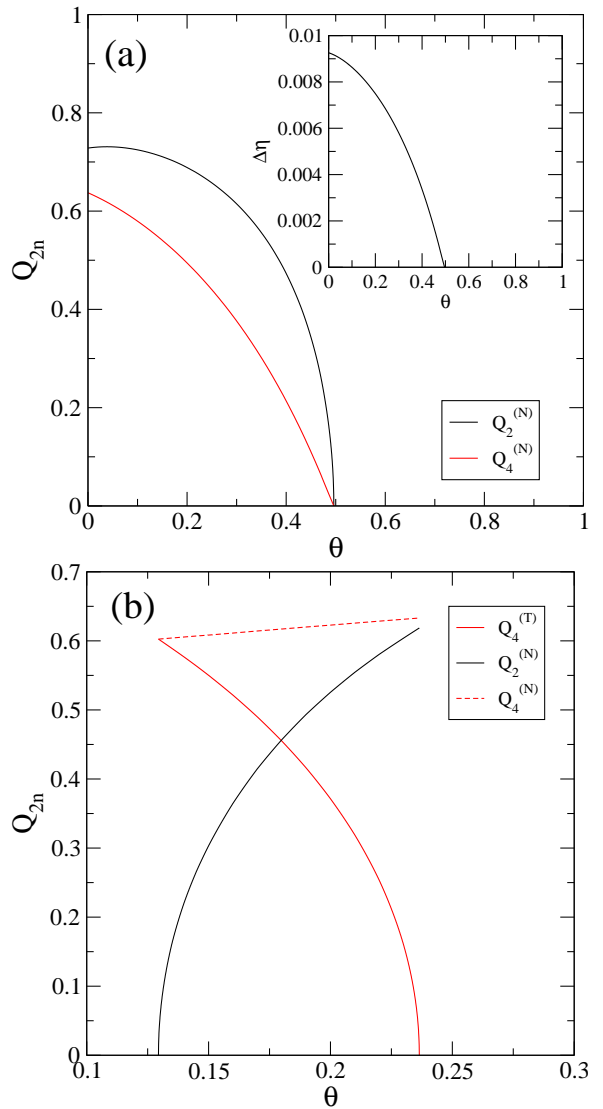

    \centering
    \epsfig{file=fig6a.eps,width=3.in}
    \epsfig{file=fig6b.eps,width=3.in}
    \caption{(a) Order parameters $Q_2$ and $Q_4$ of the coexisting N phase along the I-N coexistence of HRR with fixed 
    core length $L=2.5$ and total width $\sigma+l=1$ (implying $\theta=l$) as a function of $\theta$. Inset: Packing fraction difference 
    between the I and N phases along coexistence. (b) The same order parameters but this time along the T-N coexistence of HRR with $L=1.7$ and same total width $\sigma+l=1$.}
    \label{fig6}
\end{figure}

\begin{figure}
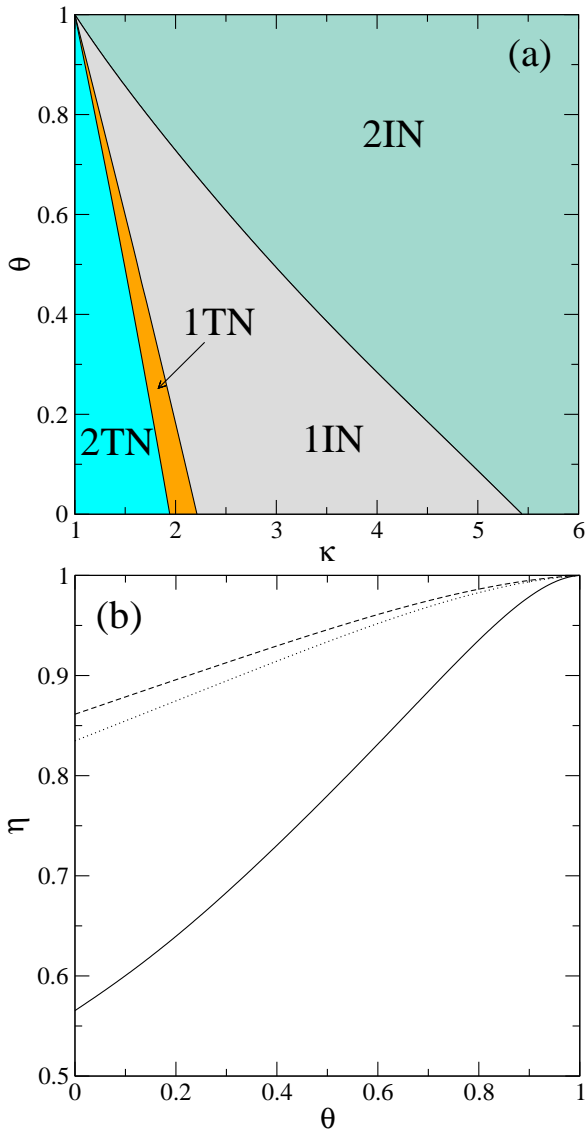

    \centering
    \epsfig{file=fig7a.eps,width=3.in}
    \epsfig{file=fig7b.eps,width=3.in}
    \caption{(a): Regions spanned by the roundness, $\theta$, 
    and aspect ratio $\kappa$ of RHR where the I-N transition is of first (1IN) or second (2IN) order and where the T-N transition is 
    of first (1TN) and second (2TN) order. (b): Packing fraction, $\eta$, at I-N (solid) and T-N (dashed) tricritical points as a function of $\theta$. With dotted line we show the value of $\eta$ corresponding to the end-critical point. }
    \label{fig7}
\end{figure}

It is important to quantify the topological changes in the phase diagrams of HRR when the roundness $\theta$ is changed. A possible way to achieve this is to calculate how the aspect ratios at the tricritical and critical end-points change as a function of $\theta$. With this information we can trace out the boundaries where first- and second-order transitions take place. To find these multi-critical points we implemented the following procedure: (i) we fixed the core length $L$ and the total width of the particle to unity: $\sigma+l=1$, implying a roundness $\theta=l$ and an aspect ratio $\kappa=L+\theta$. Next we changed $\theta$ from an initial value where a stable I-N or T-N coexistence exists, and move along the coexistence curves in the direction where the transition weakens, eventually ending in the tricritical and critical end-points. By computing the order parameters $Q_2$ and $Q_4$ as a function of $\theta$, and extrapolating their values to zero, we approximately obtain the locations of these points. In Fig. \ref{fig6} two examples of this procedure are shown. In panel (a) the evolution of $Q_2$ and $Q_4$ along the I-N transition is plotted for particles with $L=2.5$, from the initial value $\theta=0$ up to $\theta\approx 0.5$. Beyond this point the transition becomes of second order, and the I-N tricritical point can be identified. In (b) we selected $L=1.7$ and and the initial value $\theta=0.17$. Moving to the right along the T-N coexistence, and extrapolating $Q_4$ of the T phase to zero, we obtain the value of $\theta$ corresponding to the critical end-point beyond which the T-N transition turns into the I-N transition. Now going to the left, and extrapolating $Q_2$ of the N phase to zero, we obtain the T-N tricritical point beyond which the T-N transition becomes of second order. Repeating this procedure for different values of the core length $L$, one can trace out the location of three lines in the $\theta$-$\kappa$ plane: two lines correspond to the I-N and T-N tricritical points, and the third is identified with critical end-points. These lines are the boundaries of the regions where first- and second-order I-N and T-N transitions can be found. The result is shown in Fig. \ref{fig7}(a). As $\theta$ increases the range of aspect ratios where first-order transitions occur considerably shrinks, disappearing in the limit of hard discorectangles, $\theta\to 1$. 
In panel (b) the packing fractions along the multi-critical curves are represented. 
Clearly their packing fraction values increase dramatically with $\theta$ implying that, for high enough values of $\theta$, the regions of first-order transitions become unstable with respect to crystallization of the liquid-crystal fluid.  

These figures also show the loss of stability of the T phase with $\theta$: Note that in panel (a) this phase is present in the regions labeled as 2TN and 1TN, both of which considerably shrink with $\theta$. Also in panel (b) it is apparent that the packing fraction values beyond which the T phase is stable (the location of the critical end-points in dashed line) increases with $\theta$, confirming the destabilizing effect of roundness on the T phase.  

\section{Polydisperse HRR}
\label{polydisperse_sec}

 \begin{figure}
    \centering
    \epsfig{file=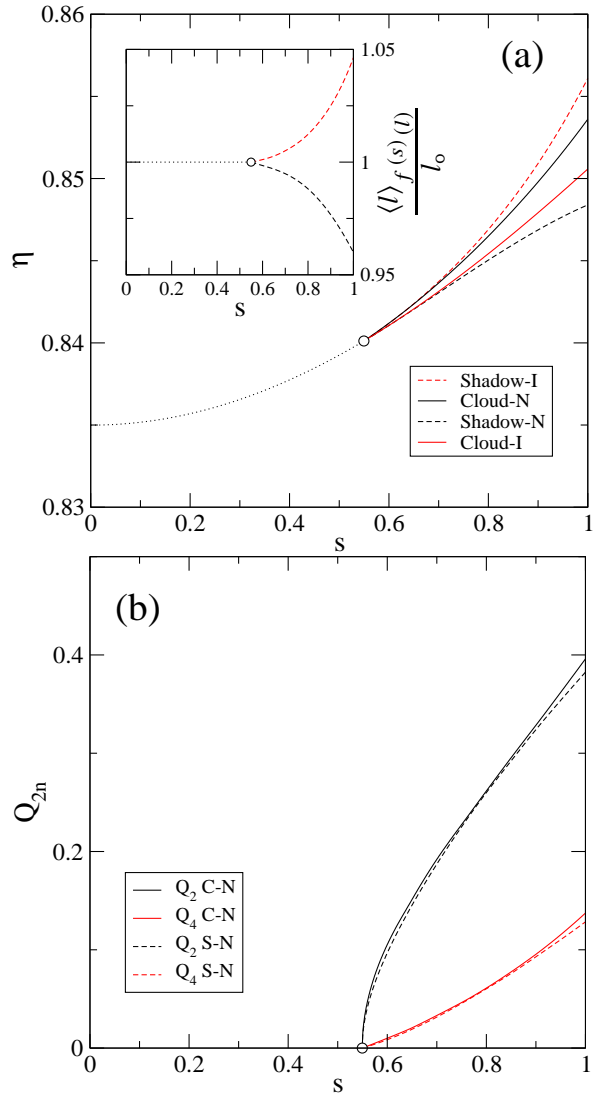,width=3.in}
    \epsfig{file=fig8b.eps,width=3.in}
    \caption{(a) Packing fractions $\eta$ vs. polydispersity coefficient $s$ for HRR with core length and width equal to $L=3$ and $\sigma=0.6$, respectively, and with mean disk diameter $l_0=1$. The resulting mean aspect ratio and roundness are $\kappa=2.5$ and $\theta=0.625$, respectively. The shadow and cloud coexistence curves (for both I and N) are correspondingly labeled. The inset shows the fractionation at the I and N shadow coexisting phases measured through the quantity $\langle l\rangle_{f^{(s)}(l)}/l_0$ (see text for definition) as a function of $s$. (b) Order parameters $Q_2$ and $Q_4$ as a function of polydispersity along the cloud and shadow coexistence curves.}
    \label{fig8}
\end{figure}

In this section we study the effect of roundness polydispersity on the phase transitions of HRR. It is first shown that, when the mean roundness $\theta$ and the polydispersity $s$ are high, the I-N transition can be of first order. Note that this transition is of second order in the one-component fluid. First we solved the set of Eqns. (\ref{looking})-(\ref{looking3}), together with the pressure equality of the coexisting phases, to find the shadow and cloud I and N coexisting curves for those values of $s$ where the I-N transition is of first order. Also, we used Eqn. (\ref{packing_I-N}) which provides an analytical expression for the packing fraction at the second-order I-N transition. We selected a particular case of core length and width with $L=3$ and $\sigma=0.6$, while the mean disk diameter is fixed to $l_0=1$. The resulting mean aspect ratio and roundness (see Eqns. (\ref{coeff})) are $\kappa=2.5$ and $\theta=0.625$, respectively. The polydisperse coefficient $s$ was varied from zero to unity, and the packing fraction at the I-N bifurcation or two-phase coexistence was calculated.

The results are plotted in Fig. \ref{fig8} (a). The transition is of second order up to a polydispersity of $s\simeq 0.55$, which is a tricritical point. Beyond this point the I-N transition is of first order, with the cloud-I and cloud-N curves inside their shadow counterparts. Note also how the shadow-N (cloud-I) curve has a lower (higher) value of $\eta$ than the cloud-I (shadow-N) curve. This means that the I phase is always enriched in particles with larger roundness than in the N phase, i.e. there is fractionation in the two-phase coexistence. Since the packing fraction depends on the mean values of $l$ and $l^2$ through the moments $m_1$ and $m_2$ (see Eqn. (\ref{def_packing}), the phase with the largest roundness will have a higher packing fraction. As a result, the orientationally ordered N phase is populated by particles with a higher mean aspect ratio $\kappa$. This effect can be better visualized in the inset of panel (a), where the mean disk diameter, scaled with $l_0$, is plotted. The average is taken with respect to the shadow-I or shadow-N length distribution functions $f^{(s)}(l)$. The inset confirms the enhancement of fractionation with $s$: the mean disk diameter is an increasing (decreasing) function of $s$ along the shadow-I (shadow-N) coexisting curve. We should bear in mind that the scaled mean diameter along the cloud coexisting curves is always equal to unity. The orientational ordering along coexistence is shown in panel (b). The order parameter $Q_2$ departs from zero at the tricritical point and follows the usual square-root law, $Q_2 \sim a\sqrt{s-s_c}$, in the neighbourhood of $s_c$. By contrast, $Q_4$ follows a linear trend. It is interesting to note how both, cloud-N and shadow-N coexisting phases, have similar orientational orderings, with the cloud phase having a slightly higher values of $Q_{2n}$ than the shadow phase, in particular for $s\sim 1$. This property of invariance in the orientational ordering is related to the fixed aspect ratio of the core, $L/\sigma$, even though the roundness polydispersity is varied: note that the sine and cosine terms in the excluded area given by Eqn. (\ref{first}), are weighted only by $L$ and $\sigma$.

\begin{figure}
    \centering
    \epsfig{file=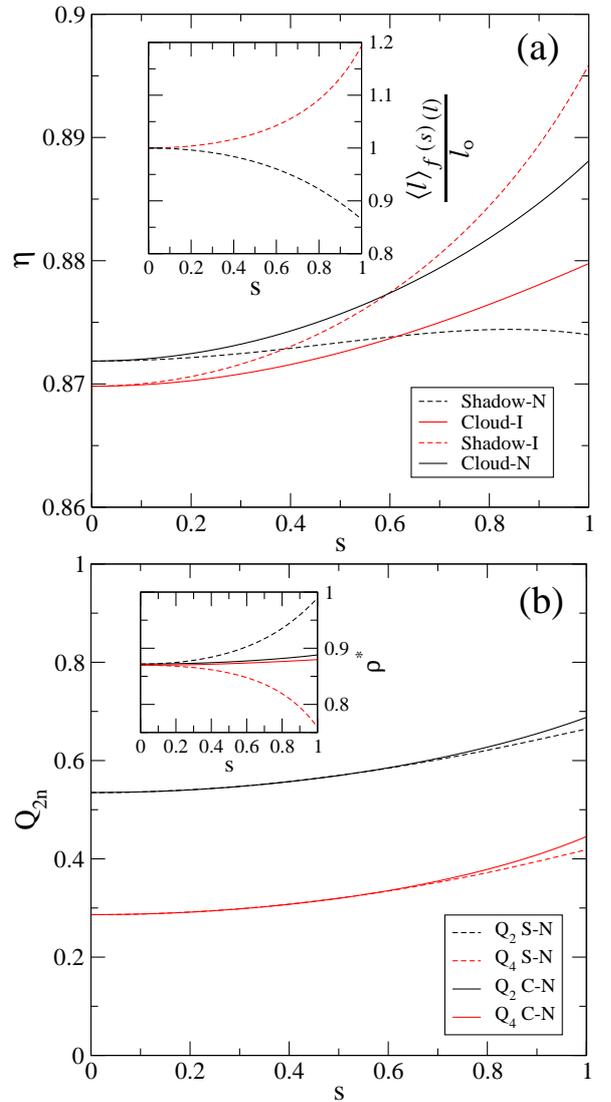,width=3.in}
    \epsfig{file=fig9b.eps,width=3.in}
    \caption{(a) Packing fraction $\eta$ vs. polydispersity $s$ for HRR with core length and width equal to $L=3$ and $\sigma=0.8$ respectively, and with a mean disk diameter $l_0=1$ (giving mean aspect ratio and roundness $\kappa=2.\overline{2}$ and $\theta=0.\overline{5}$ respectively. The shadow and cloud (for both I and N) coexistence curves are correspondingly labeled. The inset show the fractionation at the I and N shadow coexisting phases measured through the quantity $\langle l\rangle_{f^{(s)}(l)}/l_0$ as a function of s. (b) Order parameters $Q_2$ and $Q_4$ as a function of polydispersity along the cloud and shadow coexisting curves. Inset: Coexistence scaled densities $\rho^*=\rho\langle a\rangle$ as a function of polydispersity $s$ corresponding to the cloud-shadow equilibrium for both I and N phases; the labeling is the same as that of panel (a).}
    \label{fig9}
\end{figure}

\begin{figure*}
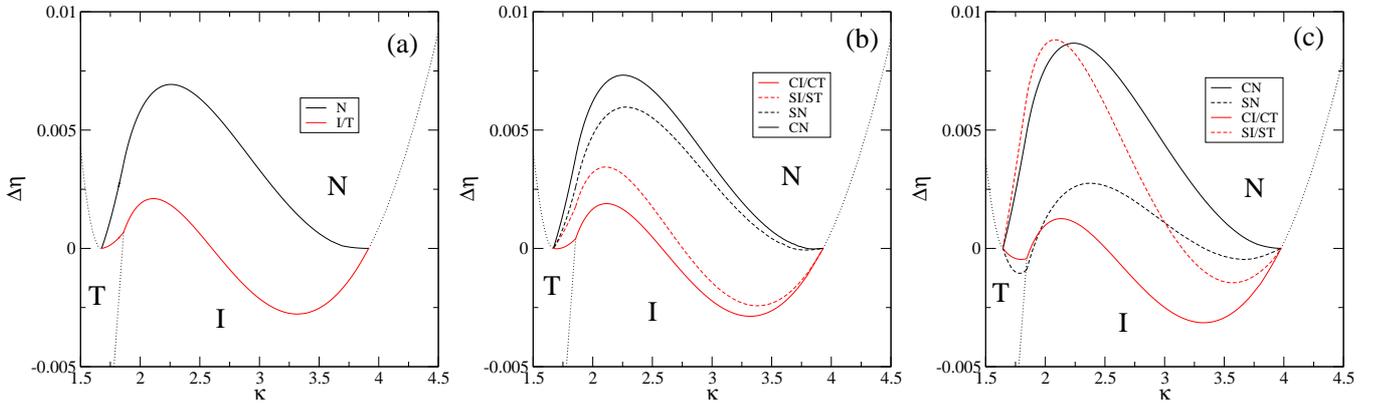

    \centering
    \epsfig{file=fig10a.eps,width=2.3in}
    \epsfig{file=fig10b.eps,width=2.3in}
    \epsfig{file=fig10c.eps,width=2.3in}
    \caption{Phase diagrams of HRR with $\theta=0.3$ in the plane $\Delta \eta$-$\kappa$, where $\Delta\eta=\eta-\eta^*(\kappa)$. $\eta^*(\kappa)=a \eta+b$ ($a$, $b$ are constants) is the straight line joining the T-N and I-N tricritical points. The polydispersity coefficients are (a) $s=0$, (b) $s=0.5$, and (c) $s=1$. Different coexistence curves and regions of stability of different phases are labeled. Dotted lines correspond to second-order phase transitions.}
    \label{fig10}
\end{figure*}

\begin{table*}
\begin{tabular}{||c|c|c|c|c||}
\hline\hline
     $\theta$ & 0 & 0.3 & 0.3 & 0.3\\
     \hline
     $s$ & 0 & 0 & 0.5 & 1\\
     \hline
     \ $\kappa_t^{\rm (TN)}$ \ & \ 1.940 \ & \ 1.676 \  & \ 1.669 \ & \ 1.644 \ \\
     \hline
     $\kappa_{ec}$ & 2.210 & 1.861 & 1.857 & 1.844\\
     \hline
     $\kappa_t^{\rm (IN)}$ &  5.440 & 3.915 & 3.932 & 3.976\\
     \hline\hline
\end{tabular}
\caption{Tabulated values of the mean aspect ratios at the I-N
($\kappa_t^{(\rm IN)}$) and T-N ($\kappa_t^{(\rm TN)}$) tricritical points, and also at the critical end-point ($\kappa_{ec}$) for different values of polydispersity $s$ and fixed value of roundness, $\theta=0.3$. The corresponding values for HR ($\theta=0$) are included.}
\label{table}
\end{table*}

\begin{figure}
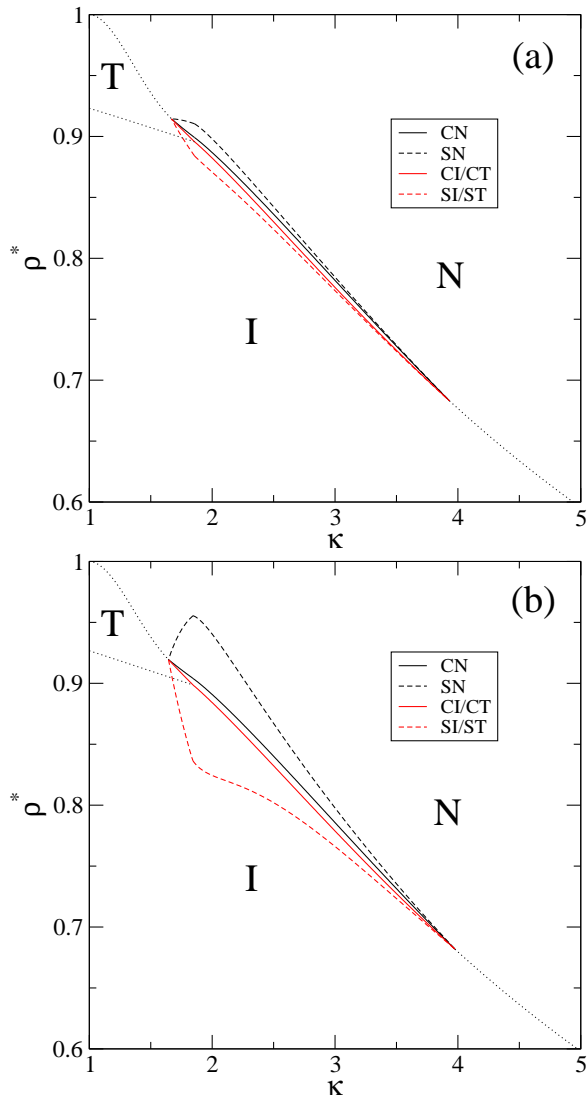

    \centering
    \epsfig{file=fig11a.eps,width=3.in}
    \epsfig{file=fig11b.eps,width=3.in}
    \caption{Phase diagrams of HRR with $\theta=0.3$ in the plane $\rho^*-\kappa$, where $\rho^*=m_0\langle a\rangle$ is the scaled number density. The polydispersity coefficients are 
    (a) $s=0.5$, and (b) $s=1$. Different coexistence curves and regions of stability of different phases are labeled. Dotted lines correspond to second-order phase transitions.}
    \label{fig11}
\end{figure}

Next we describe the changes in the phase behavior of a fluid, whose one-component counterpart exhibits a first-order I-N phase transition, when polydispersity is switched on. The particle geometry was chosen such that length and width are $L=3$ and $\sigma=0.8$, respectively, while the mean disk diameter is $l_0=1$. The mean aspect ratio and roundness result in $\kappa=2.\overline{2}$ and $\theta=0.\overline{5}$, respectively. Coexistence packing fractions are plotted as a function of $s$ in Fig. \ref{fig9}(a). In the limit of zero polydispersity, the shadow and cloud (I or N) phases coincide, as it should be. Note how the shadow-I and shadow-N curves cross each other at $s\simeq 0.4$, and also they cross the cloud-N and cloud-I curves, respectively, at $s\simeq 0.6$. As was pointed out before, this behavior is a direct consequence of the strong fractionation effect whereby the shadow-I phase is enriched in particles with larger roundness. It is also clear from the figure that the coexistence gap is enlarged so that the first-order transition becomes stronger. In the inset of \ref{fig9} (a) the scaled mean disk diameter, averaged with respect to both I and N shadow coexisting phases, is again plotted as a function of $s$. Clearly, fractionation is strongly enhanced by polydispersity. The inversion in packing fraction of the coexisting I and N phases described above is not visible in the number density, as shown in the inset of panel (b) where the properly scaled coexistence densities are plotted as a function of polydispersity. Densities display the usual behavior: the shadow-I (shadow-N) coexistence phase has a lower (higher) density than the cloud-N (cloud-I) phase. Therefore, the inversion in packing fraction is not related with a concentration effect but, as explained above, with the strong fractionation. Finally, panel (b) shows the coexistence values of the order parameters $Q_{\rm 2n}$ as a function of polydispersity. They both increase monotonically with $s$. Cloud and shadow values are very similar, once more a consequence of the invariant orientational ordering of particles with a fixed core. 

All of the above results pertain to the effect of polydispersity on the phase behavior of HRR when the core dimensions $L$ and $\sigma$ are both fixed. Now we describe how the whole phase diagram in the plane $\eta-\kappa$ evolves with polydispersity. Since $\kappa$ is varied, $L$ and/or $\sigma$ will change. We fixed a moderate mean roundness parameter $\theta=0.3$ and calculated three different phase diagrams: (i) that corresponding to the one-component fluid with no polydispersity (already described in Sec. \ref{one-component}), (ii) that with polydispersity coefficient $s=0.5$ and (iii) that with $s=1$. 

The results are shown in Fig. \ref{fig10}. Instead of the packing fraction $\eta$, the 
difference between the packing fraction and the straight line $\eta^*=a\kappa +b$ connecting the T-N and the I-N tricritical points is used for better visualizing the cloud and shadow curves, and the coexisting gap. A first result is that, for high enough polydispersity, the coexistence gap is enlarged with respect to the one-component case. Also, for $s=0.5$, both shadow I and N curves are inside the cloud curves, with the packing fraction of the coexisting I phase (shadow or cloud) being below that of the coexisting N for any aspect ratio. This is the usual trend. However, for the maximum polydispersity ($s=1$) the shadow-I and shadow-N curves intersect at $\kappa\approx 3$, and they also cross their respective cloud curves at $\kappa\approx 2$ and 2.2, a consequence of the fractionation effect. It is interesting to note that the T-N coexistence also exhibits strong fractionation, with the T phase enriched in particles with higher roundness and lower aspect ratio, while the opposite occurs with the N phase. Note how the I (shadow or cloud) coexistence curves are always above their N (cloud or shadow) counterparts in the region of T-N coexistence. When the mean roundness is moderate, as in the case $\theta=0.3$, 
the position of the I-N and T-N tricritical and critical end-points do not change appreciably with polydispersity, see Table \ref{table}. Although the general trend is that 
polydispersity enlarges the distance between the tricritical points (the I-N point moving to the right and the T-N point moving to the right), the difference is visible only in the second decimal (see table). Plotting phase diagrams in the scaled density-aspect ratio plane, it is apparent that the shadow and cloud curves follow the usual trend, with the I or T phases having lower 
densities compared with the density of N phase (see Fig. \ref{fig11}). Also, the coexistence gap becomes considerably larger with polydispersity. 

\begin{figure}
    \epsfig{file=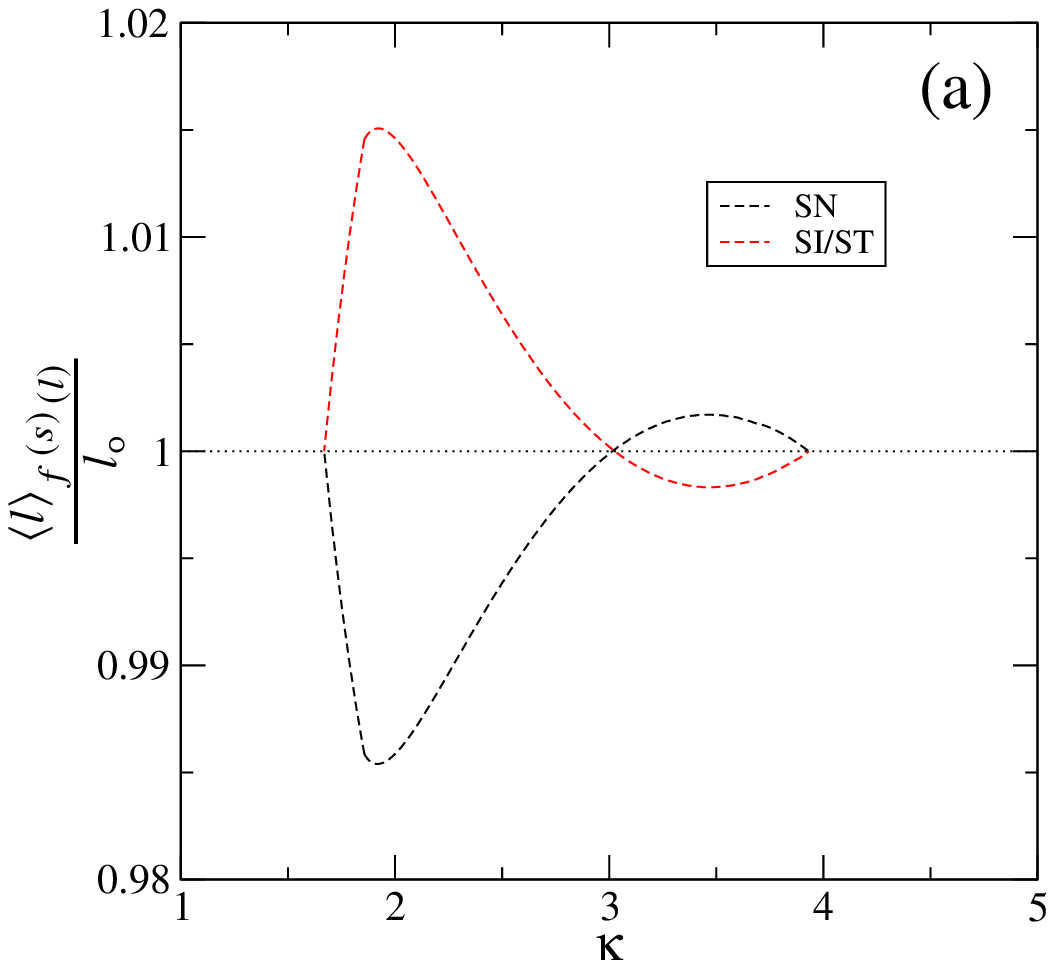,width=3.in}
    \epsfig{file=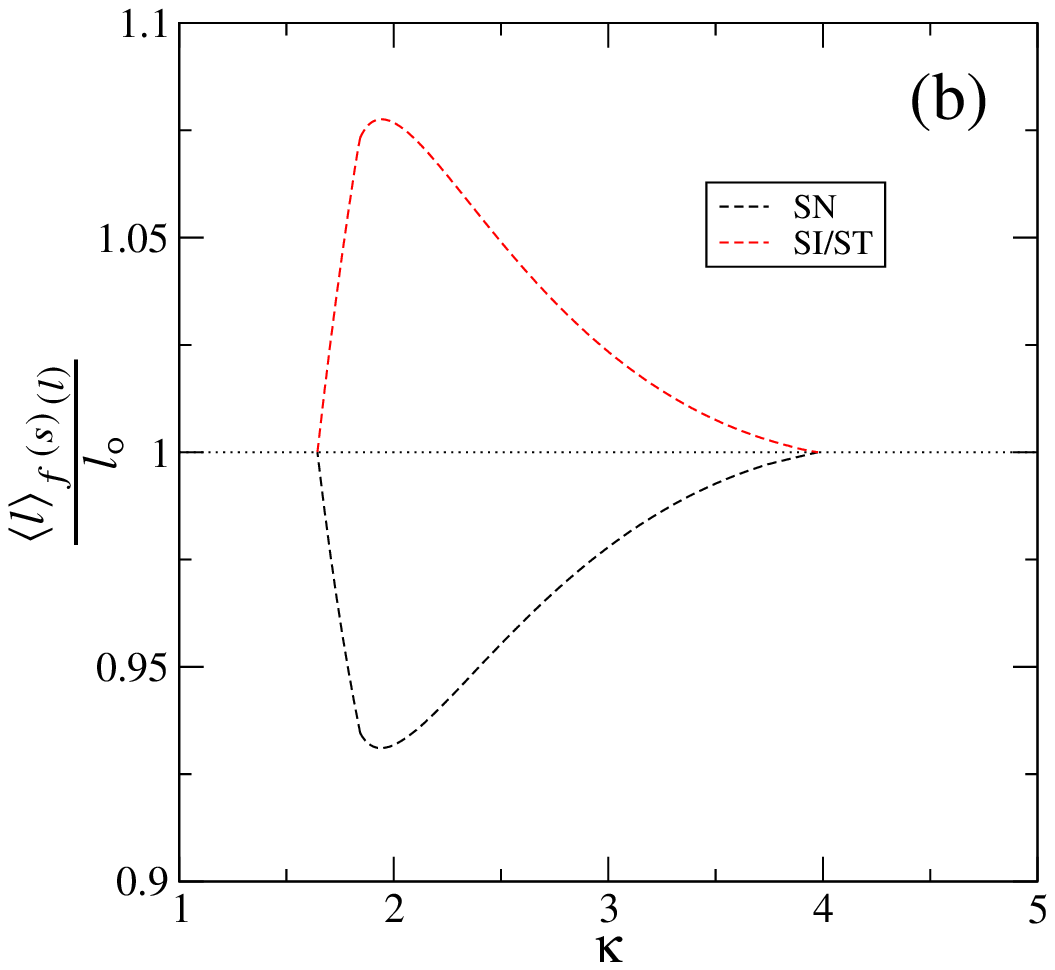,width=3.in}
    \caption{Fractionation vs. $\kappa$ measured through the quantity $\langle l\rangle_{f^{(s)}(l)}/l_0$ along the shadow coexistence curves of HRR (see keybox), with $\theta=0.3$ and polydispersity coefficients (a) $s=0.5$ and (b) $s=1$.}
    \label{fig12}
\end{figure}

To end this section, Fig. \ref{fig12} quantifies the fractionation between the coexisting phases by showing the averaged disk diameter (with respect to the shadow distribution functions), scaled with $l_0$, as a function of $\kappa$ and along the coexistence binodals. As expected, the shadow-I phase has a large proportion of particles with high roundness as compared to the cloud-N phase (both cloud phases has this magnitude fixed to 1). The opposite trend is exhibited by the shadow-N, which is enriched in less rounded particles. This general trend holds for both polydispersities, $s=0.5$ and $s=1$, except at a relatively small interval of aspect ratios, $3<\kappa<4$, corresponding to the case $s=0.5$, see panel (a), for which the I-N phase transition is relatively weak

\section{Crystalline ordering}
\label{crystal}

In the present study we have not taken into account the stability of nonuniform phases, but they certainly should be present at high enough densities, at least in the one-component fluid. In Ref. \cite{Escobedo} the authors study the effect of roundness of hard squares on the stability of the T phase and find that, if the roundness is larger than 0.3 (a value used here to predict some phase diagrams), the one-component fluid will exhibit a direct transition from the isotropic to a crystalline phase. We can use a simple argument to extend this result to HRR and estimate the critical roundness parameter beyond which a crystal phase is not expected for rounded rectangles (note that an added polydispersity, not contemplated in this section, would tend to destabilize the crystal phase regardless of the value of roundness).

To estimate this maximum roundness, $\theta_{\rm max}$, at which the one-component HRR-fluid with particles of a given aspect ratio, $\kappa>1$, destabilizes with respect to the appearance of the crystalline ordering, we use the following procedure. For $\theta_{\rm max}\approx 0.3$, the destabilization value corresponding to rounded hard squares \cite{Escobedo}, the difference between the excluded area (scaled with particle area) of hard rounded squares in parallel ($\Delta\phi=0$) or perpendicular ($\Delta\phi=\pi/2$) configurations (note that they are identical) and its maximum value at $\Delta \phi_{\rm max}$ is equal to one half of the scaled difference but at zero roundness. For rectangles we use the same criterion, extended to any aspect ratio, and write 
\begin{eqnarray}
	&&\frac{A_{\rm excl}(\Delta\phi_{\rm max},\kappa,\theta_{\rm max})-A_{\rm excl}(\pi/2,\kappa,0)}{a(\kappa,\theta_{\rm max})}\nonumber\\
	&&=\frac{A_{\rm excl}(\Delta\phi_{\rm max},\kappa,0)
    -A_{\rm excl}(\pi/2,\kappa,0)}{2a(\kappa,0)}.
    \label{ultima}
\end{eqnarray}
This equality allows to find the maximum roundness, $\theta_{\rm max}(\kappa)$, as a function of $\kappa$ for $1<\kappa\leq \kappa_c(\theta_{\rm max})$. We define $\kappa_c(\theta_{\rm max})=\kappa^*-(\kappa^*-1)\theta_{\rm max}$ as the aspect ratio corresponding to the intersection between the I-N and T-N spinodals (beyond which the T phase is no longer stable). Note that, in Eqn. (\ref{ultima}) we are using the excluded area evaluated at the T-like configuration, i.e. for $\Delta\phi=\pi/2$. 

\begin{figure}
        \centering
        \epsfig{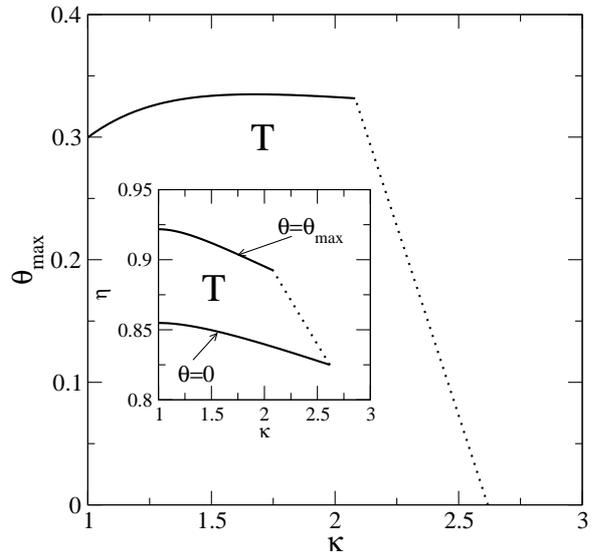}
        \caption{The function $\theta_{\rm max}(\kappa)$ (see text) and the straight line $\kappa_c=\kappa^*-(\kappa^*-1)\theta$ for $0\leq\theta\leq \theta_{\rm max}$. Both curves approximately enclose the region where the T phase of HRR is stable. Inset: packing fraction $\eta$ at the I-T bifurcation, evaluated at $\theta=0$ and 
        $\theta=\theta_{\rm max}$ (continuous curves), and packing fraction at the intersection between the I-T and I-N spinodals (dotted line), both as a function of aspect ratio $\kappa$.}
        \label{fig13}
    \end{figure}

The function $\theta_{\rm max}(\kappa)$ is plotted in Fig. \ref{fig13}. The dotted line is the straight line $\kappa_c=\kappa^*-(\kappa^*-1)\theta$ for $0\leq\theta\leq \theta_{\rm max}$. We can see that the maximum roundness, although slightly higher, has approximately the same value, for any aspect ratio,  as for hard squares. The curves shown in the inset of Fig. \ref{fig13} correspond to the packing fractions of the I-T bifurcation, evaluated at $\theta=0$ and $\theta=\theta_{\rm max}$. Also plotted (dotted line) is the packing fraction at the intersection between I-T and I-N spinodals. The region between these curves enclose the region of T phase stability as $\theta$ varies from 0 to its maximum value when the crystal phase preempts the T phase. Of course, this result would be the one resulting from any DFT whose uniform-density limit gives the SPT. As is well known, this theory overestimates the packing fraction at which liquid-crystal and nonuniform phases begin to be stable, especially for small aspect ratios. More sophisticated theories, with the inclusion of three-body of higher correlations \cite{three-body}, are necessary to describe quantitatively the phase behavior at these aspect ratios. However we are confident that the qualitative description (except for the precise packing fraction location) of the phase behavior of HRR described in the present study is the correct one.  
    
\section{Conclusions}
\label{conclusions}
In this paper we have defined a new particle model, hard rounded rectangles, to study the effect of roundness on the stability of the T phase and the character (first vs. second order) of the phase transitions involved in the phase behavior of the fluid. The first part of the study is devoted to characterizing the changes in the phase diagram of the one-component fluid, while in the second part a continuous polydispersity in the roundness parameter at fixed core lengths is introduced in order to identify novel trends in the phase behavior of the polydisperse fluid.

For the one-component fluid we have found that the main effect of roundness is the destabilization of the T phase: the stability region considerably shrinks in aspect ratio, and moves to higher densities. Also the I-N and T-N transitions, which are of first order in the hard-rectangle fluid, transform into second order or become weaker, the latter scenario occurring if the aspect ratio lies inside the interval defined by the T-N and I-N tricritical points. This interval dramatically shrinks with roundness parameter, disappearing altogether in the limit of discorectangles. In addition, we believe that the crystal phase should not interfere with the above scenario as our estimations lead to a wide range of values for the roundness parameter where the T phase should be stable against the crystal.

When polydispersity is added, and if its magnitude and the mean roundness are large enough, the I-N transition for certain mean aspect ratios changes from second to first order, with the presence of fractionation in roundness between the coexisting phases. We quantified this fractionation by measuring the mean disk diameter at both coexisting phases, resulting in a N phase enriched in particles with low roundness. Another important effect when polydispersity is large enough is a packing-fraction inversion: when the I-N or T-N transitions are of first order, the I and T phases have higher packing fraction than the N phase. This phenomenon is related to fractionation: due to the enrichment of the I or T phases in species with high roundness when they coexist with the N phase, the packing fraction, being a function of the first and second moments of the disk-diameter distribution function, will have a larger value.

Finally, as regards how the crystalline phase could modify the results presented in this study, we expect that the inclusion of a large enough polydispersity will destabilize the crystal phase with the effect of increasing the threshold value in mean roundness beyond which the crystal phase becomes stable.  

\appendix 

\section{Explicit equations for coexistence and bifurcation calculations}
\label{app}
 
The set of coexistence equations (\ref{todas}) can be simplified using the following properties of the 
function $A_{\rm spt}(\Delta\phi,l,l')$ (see Eqs. (\ref{first}) and (\ref{spt0})): 
(i) Sine an cosine functions in the relative angle $\Delta\phi$ are decoupled from the terms that contain $l$ and $l'$. This in turn 
implies that the double average of $A_{\rm spt}$ with respect to $\rho(l,\phi)$ is equivalent to 
the average of the sine and cosine functions with respect to $m_0(\phi)$ 
(see Eqn. (\ref{spt})). Using the Fourier expansion (\ref{expansion}), this
in turn gives a result which only depends on the Fourier coefficients $\{m_0^{(k)}\}$ of the zeroth moment,  
while the first moment enters only through its integrated value $m_1$. (ii) Moreover, the first functional derivative of $\langle\langle A_{\rm spt}(\Delta\phi,l,l')\rangle\rangle_{\rho(l,\phi)}$
with respect to $\rho(l,\phi)$, which it is needed to calculate the function $c_1(l,\phi)$,
will also depend on $\{m_0^{(k)}\}$ (see Eqn. (\ref{delta})), which are decoupled from the polydisperse variable $l$. 
(iii) Finally, the dependence of $c_1(l,\phi)$ on $l$ is a linear polynomial. This is very important because taking into account the expression for $f(l)$ from Eq. (\ref{parent}), 
the integration over $l$ in the coexistence set of equations 
(\ref{todas}) can be computed analytically. Defining the new variables 
$y^{(\alpha)}_i\equiv m_i^{(\alpha)}/(1-\eta_{\alpha})$ with $i=0,1,2$ and 
$\alpha=c,s$ we obtain, from (\ref{todas}) and all the above properties, the result:
\begin{eqnarray}
	&&y_i^{(s)}=y_0^{(c)} e^{-\Delta_1(\{y_j\})}\chi_i(\{y_j\})\frac{\displaystyle\int_0^{2\pi}
	d\phi e^{-\tilde{c}_1^{(s)}\left(\{y_0^{(j)}\},\phi\right)}}
	{\displaystyle\int_0^{2\pi} 
	d\phi e^{-\tilde{c}_1^{(c)}\left(\{y_0^{(j)}\},\phi\right)}}, \nonumber\\ 
	&&i=0,1,2 \label{looking}\\
	&&y_0^{(k,s)}=2y_0^{(c)} e^{-\Delta_1(\{y_j\})}\chi_0(\{y_j\}) 
	\nonumber\\
	&&\times\frac{\displaystyle\int_0^{2\pi}
	d\phi \cos(2k\phi) e^{-\tilde{c}_1^{(s)}\left(\{y_0^{(j)}\},\phi\right)}}
	{\displaystyle\int_0^{2\pi} d\phi e^{-\tilde{c}_1^{(c)}\left(\{y_0^{(j)}\},\phi\right)}},\label{looking2}\\
	&& y_0^{(k,c)}=2y_0^{(c)}
	\frac{\displaystyle\int_0^{2\pi}
	d\phi \cos(2k\phi) e^{-\tilde{c}_1^{(c)}\left(\{y_0^{(j)}\},\phi\right)}}
	{\displaystyle\int_0^{2\pi} d\phi e^{-\tilde{c}_1^{(c)}\left(\{y_0^{(j)}\},\phi\right)}}, \label{looking3}
\end{eqnarray}
while, for the cloud phase, we have $\displaystyle{y_i^{(c)}=\frac{\rho_0 \langle l^i\rangle_{f(l)}}{(1-\eta_c)}}$. 
In the preceding equations we have defined 
\begin{eqnarray}
	&&\Delta_0(\{y_i\})=\left(L+\sigma\right)\left(y_0^{(s)}
	-y_0^{(c)}\right)+\frac{\pi}{2}\left(
	y_1^{(s)}-y_1^{(c)}\right), \nonumber\\ 
	&&\\ 
	&&\Delta_1(\{y_i\})=\frac{2}{\pi}\left(L+\sigma\right)\Delta_0(\{y_i\}),
	\\
	&&\chi_i(\{y_j\})=\frac{\langle l^i\rangle_{f(l)}}
	{\left(1+\Delta_0(\{y_j\}) l_0 s^2\right)^{i+s^{-2}}},\\
	&& \tilde{c}_1^{(\alpha)}(\{y_0^{(j)}\})=-\frac{2}{\pi}
	\sum_{k\geq 1}g_k y_0^{(k,\alpha)}\cos(2k\phi), \label{lac}
\end{eqnarray}
and the packing fractions of the cloud and shadow phases can be calculated from $\{y_i^{(\alpha)}\}$ as 
\begin{eqnarray}
	\eta_{\alpha}=1-\frac{1}{1+L\sigma y_0^{(\alpha)}+\left(L+\sigma\right)
	y_1^{(\alpha)}+\pi y_2^{(\alpha)}/4}.
	\label{packing}
\end{eqnarray}
Note that, in the definition of $\tilde{c}_1^{(\alpha)}(\{y_0^{(j)}\})$, a term proportional to the pressure $p^{(\alpha)}$ does not appear (as in Eqn. (\ref{c1})) because it should be in the numerator and denominator of Eqns. (\ref{looking})--(\ref{looking3}). As both pressures should be equal at coexistence they cancel. 
The pressure within the new variables is
\begin{eqnarray}
	&&\beta p_{\alpha}=y_0^{(\alpha)}+\frac{1}{\pi}\left(\left(L+\sigma\right)
	y_0^{(\alpha)}+\frac{\pi}{2}y_1^{(\alpha)}\right)^2\nonumber\\
	&&-\frac{1}{2\pi}\sum_{k\geq 1} g_k\left(y_0^{(k,\alpha)}\right)^2.
	\label{press}
\end{eqnarray}
Looking at Eqns. (\ref{looking}) we can see that the variables 
$\{y_0^{(s)},\ y_1^{(s)}, \ y_2^{(s)}\}$ are not independent. Dividing 
Eqns. for $y_i^{(s)}$ ($i=1,2$) by that for $y_0^{(s)}$, we obtain
\begin{eqnarray}
	&&\frac{y_i^{(s)}}{y_0^{(s)}}=
	\frac{1+(i-1)s^2}{s^{2i}\left(T\left(\{y_0^{(\alpha)}\}\right)
	+\pi y_1^{(s)}/2\right)^i}, \label{solve}\\
	&&T\left(\{y_0^{(\alpha)}\}\right) 
	=\frac{1}{l_0 s^2}+\left(L+\sigma\right)y_0^{(s)}\nonumber\\
	&&-
	\left(L+\sigma+\frac{\pi}{2}l_0
	\right)y_0^{(c)}.
\end{eqnarray}
Eqns. (\ref{solve}) can be solved for $y_i^{(s)}$ ($i=1,2$) as a function of 
$y_0^{(s)}$ and $y_0^{(c)}$ to find 
\begin{eqnarray}
	&&y_2^{(s)}=\frac{(1+s^2)}{y_0^{s}} \left(y_1^{(s)}\right)^2,\\
	&&y_1^{(s)}=\frac{1}{\pi}\left[\sqrt{T\left(\{y_0^{(\alpha)}\}\right)^2
	+\frac{2\pi y_0^{(c)}}{s^2}}-T\left(\{y_0^{(\alpha)}\}\right)\right]
	\label{obtain}\nonumber\\
\end{eqnarray}
Finally the function $\chi_0(\{y_i\})$ can be computed as 
\begin{eqnarray}
	\chi_0(\{y_i\})=\left(\frac{y_1^{(s)}}{y_0^{(s)}l_0}\right)^{s^{-2}},
\end{eqnarray}
which taking into account (\ref{obtain}) is also a function of $\{y_0^{(s)},\ y_0^{(c)}\}$. 

Taking into account the preceding discussion, we have a total number of $2(N_{\rm max}+1)$ independent variables 
$\{y_0^{(c)},\ y_0^{(s)},\ y_0^{(k,c)},\ y_0^{(k,s)}\}$, where $N_{\rm max}$ is the total number of Fourier amplitudes used 
in the truncated Fourier expansion (\ref{expansion}). Thus we need to solve the single Eqn. (\ref{looking}) for $y_0^{(s)}$ 
($i=0$) and Eqns. (\ref{looking2}) and  (\ref{looking3}) for the total number of $2N_{\rm max}$ Fourier amplitudes $y_0^{(k,\alpha)}$ 
of the cloud ($\alpha=c$) and shadow ($\alpha=s$) phases. Finally the unknown $y_0^{(c)}$ can be computed 
from the equality of pressures, $p_c=p_s$, 
which guarantees mechanical equilibrium between the coexisting phases. From this equality and Eqn. (\ref{press}) 
we can see that 
the variable $y_0^{(c)}$ can be written as a function of $y_0^{(s)}$ and $y_0^{(k,\alpha)}$ by 
solving a quadratic equation. Note that the
$2(N_{\rm max}+1)$ variables correspond to the case where both, the cloud and shadow phases have orientational ordering, 
for example when the N and T phases coexist. When one of the coexisting phases, say the cloud phase, is I we need 
to solve only $N_{\rm max}+2$ equations because $y_0^{(k,c)}=0$ for $1\leq k\leq N_{\rm max}$.
We have solved Eqns. (\ref{looking})-(\ref{looking3}) through a mixed Piccard iteration method, stopped when a prescribed tolerance criterion is achieved,
$\displaystyle\sum_{\alpha=c,s}\sum_{k=0}^{N_{\rm max}}\left|y_{0,n+1}^{(k,\alpha)}-y_{0,n}^{(k,\alpha)}\right|\leq 10^{-7}$, 
where $n$ label the number of iteration.

When a second order I-(N,T) transition takes place, we can calculate the corresponding
packing fraction as follows: In the close 
neighbourhood of the instability of the I phase with respect to 
the N or T phases we can expand Eqn. 
(\ref{looking3}) with respect to the small quantity $y_0^{(k,c)}$ 
(with $k=1$ and 2 
for N and T symmetries respectively) up to first order. Taking into account the 
expression (\ref{packing}) for the packing fraction as a function of $y_i^{(c)}$, and 
the fact that $y_i^{(c)}=y_0^{(c)}\langle l^i\rangle_{f(l)}$, with 
$\langle l\rangle_{f(l)}=l_0$ and $\langle l^2\rangle_{f(l)}=l_0^2(1+s^2)$, we
finally obtain 
\begin{eqnarray}
	\eta_k=\left(1+\frac{2g_k}{\pi\langle a\rangle_{f(l)}}\right)^{-1}. 
	\quad k=1,2,
	\label{bifurca}
\end{eqnarray}
where the mean particle area is defined as 
\begin{eqnarray}
	\langle a\rangle_{f(l)}=L\sigma +(L+\sigma)l_0+\frac{\pi}{4}l_0^2
	(1+s^2).
\end{eqnarray}

The packing fraction at bifurcation from T to N phase can be obtained by 
expanding the exponentials of Eqn. (\ref{looking3}) with respect to $y_0^{(2n-1,c)}$, 
the small odd-Fourier amplitudes,  up to first order and evaluating the 
resulting integrals at the equilibrium even-Fourier amplitudes $y_0^{(2n,c)}$ 
(which may be quite large because the T phase can have a high orientational
order). We thus obtain the following set of linear equations, written in matrix 
form $B\cdot {\bm t}={\bf 0}$ with matrix elements
\begin{eqnarray}
	&&b_{kn}=\delta_{kn}-\frac{1}{\pi} g_{2n-1}\nonumber\\
	&&\times \left(
	y_0^{(2(k+n-1),c)}+y_0^{(2|k-n|,c)}(1+\delta_{kn})\right), \label{linear}
\end{eqnarray}
with $\delta_{kn}$ the Kronecker-delta, ${\bm t}=\left(y_0^{(1,c)},\dots,y_0^{(2m-1,c)}\right)^T$ and $m=N_{\rm max}/2$. 
This system has a nontrivial solution only if ${\cal B}(y_0^{(c)})=
\text{det} \left(B\right)=0$, which allows to find $y_0^{(c)}$ (we should 
bear in mind the notation $y_0^{(0,c)}=y_0^{(c)}$) at bifurcation, 
and from this, and Eqn. (\ref{packing}), the value of packing fraction.

The numerical procedure to calculate the two-phase coexistence 
in the one-component limit can be obtained from the same Eqs. 
(\ref{looking})-(\ref{looking3}) by setting $i=0$ and taking into account the limit
\begin{eqnarray}
	&&\lim_{s\to 0} \chi_i\left(\{y_j\}\right)=l_0^i e^{-\Delta_0(\{y_j\})}, 
	\\ &&\Delta_0(\{y_j\})=\left(L+\sigma+\frac{\pi}{2}l_0\right)
	\left(y_0^{(s)}-y_0^{(c)}\right),
\end{eqnarray}
while the packing fraction at the I-(N,T) bifurcation is given 
by (\ref{bifurca}) setting $s=0$.

\acknowledgements

Financial support from Grant No. PGC2018-096606-B-I00 (MCIU/AEI/FEDER,UE) is acknowledged.

\end{document}